\documentclass[pra,twocolumn,showpacs,superscriptaddress]{revtex4}
\usepackage{amsmath}
\usepackage{amsfonts}

\begin{document}
\title{Universal Continuous--Variable Quantum Computation: \\
       Requirement of Optical Nonlinearity for Photon Counting}
\author{Stephen D.\ Bartlett}
\affiliation{Department of Physics and
        Centre for Advanced Computing -- Algorithms and Cryptography,   \\
        Macquarie University,
        Sydney, New South Wales 2109, Australia}
\author{Barry C.\ Sanders}
\affiliation{Department of Physics and
        Centre for Advanced Computing -- Algorithms and Cryptography,   \\
        Macquarie University,
        Sydney, New South Wales 2109, Australia}
\affiliation{Quantum Entanglement Project, ICORP, JST,
        Edward L.\ Ginzton Laboratory, \\ Stanford University, 
        California 94305-4085} 
\date{March 19, 2001}

\begin{abstract}
  Although universal continuous--variable quantum computation cannot
  be achieved via linear optics (including squeezing), homodyne
  detection and feed--forward, inclusion of ideal photon counting
  measurements overcomes this obstacle.  These measurements are
  sometimes described by arrays of beamsplitters to distribute the
  photons across several modes.  We show that such a scheme cannot be
  used to implement ideal photon counting and that such measurements
  necessarily involve nonlinear evolution. However, this requirement
  of nonlinearity can be moved `offline', thereby permitting universal
  continuous--variable quantum computation with linear optics.
\end{abstract}
\maketitle

\section{Introduction}

The chief attraction of quantum computation is the possibility of
solving certain problems exponentially faster than any known method on
a classical computer~\cite{Sho94}, and a significant effort is
underway to realize a physical quantum computer~\cite{Ste98,Nie00}.
Optical realizations of a quantum computer are particularly appealing
because of the robust nature of quantum states of light against the
effects of decoherence as well as the advanced techniques for state
preparation, photon manipulation, and photodetection.  Both
discrete--variable (qubit--based)~\cite{Chu95,Kni01,Got01b} and
continuous--variable (CV)~\cite{Llo99} schemes offer significant
potential as optical quantum computers.  However, the lack of a strong
optical nonlinearity has been a considerable hurdle for optical
quantum computation~\cite{Nie00}.

A proposal by Knill, Laflamme, and Milburn (KLM)~\cite{Kni01}
describes how measurement of photons can be employed to induce a
nonlinear transformation in a qubit--based optical quantum computer
and how this procedure can be done efficiently in a non--deterministic
way.  These remarkable results suggest that measurement in a CV system
may be used to induce nonlinear evolution as well (although
measurements of CV observables may not be possible in the von Neumann
sense; see~\cite{Oza97}).  A scheme proposed by Gottesman, Kitaev, and
Preskill (GKP)~\cite{Got01b} uses photon number measurement to induce
a nonlinear transformation.  If photon counting overcomes the obstacle
of creating optical nonlinearities, then CV quantum computation may be
feasible, just as the qubit--based linear optical quantum computer may
be feasible~\cite{Kni01}.  Here we present three key results:
\begin{enumerate}
\item universal quantum computation over continuous variables can be
  achieved using linear optics, homodyne measurement with
  feed--forward, and photon counting,
\item the desired photon counting projective measurement cannot be
  performed using linear optics and existing photodetectors, and
  necessarily involves an optical nonlinearity, and
\item the nonlinear transformations can be brought `offline' to
  prepare quantum resources for a linear optical CV quantum
  computation to succeed in a deterministic way.
\end{enumerate}

This paper is outlined as follows.  We first review the Clifford group
for continuous variables, which consists of linear optics
transformations (including squeezing).  The construction of a cubic
phase state of GKP is outlined as a possible means to implement a
nonlinear transformation, necessary for universal CV quantum
computation.  This gate requires photon counting measurements, and we
demonstrate that linear optics, homodyne measurement, and ideal
photodetection is insufficient to implement such schemes.  A nonlinear
interaction is shown to be a necessary component of any photon
counting measurement.  An analysis of truncated Hilbert spaces is
included, and the paper concludes with a discussion on the use of
photon counting measurements (and their associated nonlinear
transformations) offline in a CV quantum computation.

\section{Clifford group transformations}

The requirement of nonlinear transformations for universal CV quantum
computation~\cite{Llo99} can be understood by considering $n$ harmonic
oscillators, corresponding to $n$ independent optical field modes,
with annihilation operators $\{ \hat{a}_i; i=1,\ldots,n\}$.  Linear
optical transformations of these modes are described by unitary
phase--space displacements (by mixing with `classical fields' at beam
splitters) 
\begin{equation}
  \label{eq:Displacements}
  D_i(\alpha) = \exp(\alpha \hat{a}_i^\dag - \alpha^* \hat{a}_i ) \, ,
\end{equation}
with $\alpha \in \mathbb{C}$; these transformations comprise
the Heisenberg--Weyl group HW$(n)$.  For a classical pump field,
parametric amplification invokes one--mode squeezing operations
\begin{equation}
  \label{eq:OneModeSqueezing}
  S_i(\eta) = \exp(\tfrac{1}{2}(\eta \hat{a}_i^{\dag\,2} - \eta^*
  \hat{a}_i^2) ) \, ,
\end{equation}
and two--mode squeezing operations 
\begin{equation}
  \label{eq:TwoModeSqueezing}
  S_{ij}(\eta) =
  \exp(\tfrac{1}{2}(\eta \hat{a}_i^\dag \hat{a}_j^\dag - \eta^*
  \hat{a}_i \hat{a}_j )) \, ,
\end{equation}
for $\eta \in \mathbb{C}$~\cite{Cav91}.  (Although squeezing utilizes
an optical nonlinearity of order two or higher, the transformation is
regarded as being linear because the resultant Heisenberg operator
equations of motion are linear.)  Squeezing operations (both one-- and
two--mode) generate the symplectic group Sp$(2n,\mathbb{R})$.

The squeezing operation $S_i(\eta)$ with $\eta$ real maps the canonical
position as
\begin{equation}
  \label{eq:Squeezingq}
  S_i(\eta): \hat{q}_i=\sqrt{\hbar/2}(\hat{a}_i +\hat{a}_i^\dag)
  \to \exp(-\eta)\hat{q}_i\, ;
\end{equation}
thus, the infinitely squeezed displaced vacuum
\begin{equation}
  \label{eq:SqueezedVac}
  \lim_{\eta\to\infty} S_i(\eta)D_i(q/\sqrt{2\hbar})|0\rangle \, ,
\end{equation}
with $q \in \mathbb{R}$ is the (unnormalizable) position
eigenstate $|q\rangle_i$.  These position eigenstates are often
employed as a computational basis for CV quantum computation, and are
approximated in experiment by finite squeezing~\cite{Llo99}.
Two--mode squeezing $S_{ij}(\eta)$ acts in a similar fashion on the
normal and antinormal modes $\hat{a}_i \pm \hat{a}_j$, and the
infinitely squeezed two--mode vacuum 
\begin{equation}
  \label{eq:EPR}
  |\Theta\rangle_{ij} = \lim_{\eta\to\infty}S_{ij}(\eta) |0\rangle \, ,
\end{equation}
with $\eta\in\mathbb{R}$ is the EPR state satisfying
\begin{equation}
  \label{eq:EPRCond}
  {}_{ij}\langle qq'|\Theta\rangle_{ij} = \delta(q-q') \, .
\end{equation}
Two--mode squeezing also allows us to implement a unitary SUM
gate~\cite{Got01b,Bar01b}, defined as
\begin{equation}
  \label{eq:SUMgate}
  \text{SUM}_{ij} = \exp\bigl( - \tfrac{i}{\hbar}\hat{q}_i \hat{p}_j
  \bigr) = \exp\bigl(\tfrac{1}{2}(\hat{a}^\dag_i +
  \hat{a}_i)(\hat{a}^\dag_j - \hat{a}_j) \bigr)\, .
\end{equation}
This gate acts on the computational basis of position
eigenstates according to
\begin{equation}
  \label{eq:ActionOfSUM}
  \text{SUM}_{ij}:\, |q_i\rangle_i |q_j\rangle_j \to |q_i\rangle_i
  |q_i+q_j\rangle_j \, .
\end{equation}
The $i^{\text{th}}$ mode is referred to as the control and the
$j^{\text{th}}$ mode as the target.

Phase--space displacements and squeezing together close to a
finite--dimensional group known as the Clifford
group~\cite{Got01b,Bar01b}.  For $n$ modes, the Clifford group is the
semidirect product group $[\text{Sp}(2n,\mathbb{R})]\text{HW}(n)$
generated by all Hamiltonians that are inhomogeneous quadratics in the
canonical operators $\{ \hat{a}_i, \hat{a}_i^\dag, i=1,\ldots,n\}$.
The above unitary representation of the Clifford group is a subgroup
of all unitary transformations on $n$ modes.  As such, they are
insufficient to generate arbitrary unitary transformations and thus
cannot perform universal quantum computation.  The addition of a
nonlinear operation such as that provided by the $\chi^{(3)}$, or
optical Kerr, nonlinearity~\cite{Mil83} suffices, in principle, to
perform universal CV quantum computation, but is not feasible in
quantum optical implementations due to the lack of sufficiently strong
nonlinear materials with low absorption.  However, as stressed by
Lloyd and Braunstein~\cite{Llo99}, \emph{any} nonlinear coupling on a
\emph{single mode} could allow for universal CV quantum computation,
as opposed to the qubit case where a nonlinear coupling between qubits
is required.

\section{The cubic phase gate}

One is lead to ask whether measurements can be used to induce
nonlinear evolution, following the related example for qubit--based
linear optics quantum computation~\cite{Kni01}.  First, we consider
Clifford group transformations conditioned on the results of
projective--valued measurements (PVMs) in the computational basis (i.e.,
von Neumann measurements in the basis $\{ |q\rangle, q\in \mathbb{R}
\}$).  As shown in~\cite{Bar01b}, such measurements and feed--forward
are efficiently simulatable on a classical computer and thus (under
the assumption that universal quantum computation is \emph{not}
efficiently simulatable classically) are also insufficient for
universal quantum computation.  These results also include a more
realistic computational basis of finitely--squeezed states and
realistic homodyne measurement of quadratures.  Thus, in the
following, we will consider Clifford group transformations conditioned
on homodyne measurement to be part of ``linear optics''.

Possibly, measurements in a different basis can be employed to induce
a nonlinear transformation.  Such a scheme has been proposed (in the
context of quantum computation with finite--dimensional qudits rather
than CV) by GKP using measurement of photon number.  Specifically,
these measurements are described by a PVM
\begin{equation}
  \label{eq:PhotonCountingPVM}
  \bigl\{\Pi_n = |n\rangle \langle n|,\, n=0,1,2,\ldots \bigr\} \, ,
\end{equation}
for a single oscillator, where $|n\rangle$ is the eigenstate of the
number operator $\hat{N} = \hat{a}^\dag \hat{a}$ with eigenvalue $n$.
In what follows, we refer to this PVM as the \emph{photon counting
  PVM}.  The scheme of GKP is briefly outlined in the following, and
relies on the creation of a so--called ``cubic phase state''
$|\gamma\rangle$, which is the (unnormalizable) state defined as
\begin{equation}
  \label{eq:CubicPhaseState}
  |\gamma\rangle = \int \text{d}q\, \exp(i\gamma q^3) |q\rangle \, .
\end{equation}

The cubic phase state $|\gamma\rangle$ can be prepared using
squeezing, phase space displacement, and photon counting.  Consider
the two--mode squeezed vacuum state $S_{12}(\eta)|0\rangle$, $\eta \in
\mathbb{R}$, and a large momentum displacement of the first mode, to
obtain the two--mode state 
\begin{equation}
  \label{eq:wetastate}
  |w,\eta\rangle = D_1(iw)S_{12}(\eta)|0\rangle \, ,
\end{equation}
with $w \in \mathbb{R}$.  By performing a measurement of photon number
(described by the photon counting PVM) on the first mode, a
measurement result of $n$ photons projects the second mode of the pair
into a cubic phase state $|\gamma'\rangle$ to a good approximation if
$w$ is sufficiently large (details can be found
in~\cite{Cav91,Got01b}), where $\gamma' \propto n^{-1/2}$.  This state
can be transformed into $|\gamma\rangle$ with $\gamma$ of order unity
using one--mode squeezing.

The cubic phase state can be used to implement a nonlinear
transformation on an arbitrary state $|\psi\rangle_i$ of an optical mode
as follows~\cite{Got01b}.  A SUM$_{ij}^{-1}$ gate (see
Eq.~(\ref{eq:SUMgate})) is executed with $|\psi\rangle_i$ as the control
and $|\gamma\rangle_j$ as the target.  A position measurement is
performed on the target, projecting the control into the state
\begin{align}
  \label{eq:TransformedControlState}
  |\psi'\rangle_i =&\, _j\langle q=a| \, \text{SUM}_{ij}^{-1} |\psi\rangle_i
  |\gamma\rangle_j \nonumber \\
  =& \exp\bigl(i(\hat{q}_i+a)^3 \bigr)|\psi\rangle_i \, ,
\end{align}
for a measurement outcome $a$.  Invoking the Clifford group transformation
\begin{align}
  \label{eq:U(a)}
  U(a) &= \exp\bigl( i \hat{q}_i^3 - i (\hat{q}_i+a)^3 \bigr) \nonumber \\
  &= \exp(-ia^3/4)\exp\bigl(-3ai(\hat{q}_i + a/2)^2 \bigr) \, ,
\end{align}
(which can be implemented using linear optics) on the state
$|\psi'\rangle_i$ gives a net transformation equivalent to applying
\begin{equation}
  \label{eq:Vgamma}
  V_\gamma = \exp(i\gamma \hat{q}_i^3) \, ,
\end{equation}
on $|\psi\rangle_i$.  We refer to the transformation $V_\gamma$,
implemented in this manner, as the \emph{cubic phase gate}.  This
nonlinear transformation could be used in combination with Clifford
group operations to perform universal quantum computation over
continuous variables.

What is fundamentally important about this result is that \emph{all}
transformations involved, including
\begin{enumerate}
\item the preparation of the two--mode squeezed state,
\item the transformations needed to prepare $|\gamma\rangle$,
\item the SUM$^{-1}$ gate, and
\item the transformation $U(a)$ (which is quadratic in $\hat{q}$),
\end{enumerate}
are implementable using Clifford group (linear optics)
transformations.  The resulting cubic phase gate is fully
deterministic.  The key component that allows for the nonlinear
transformation is the measurement of photon number.  In the following,
we argue that such a measurement possesses ``hidden'' nonlinear
evolution (i.e., the equivalent of an optical Kerr or higher order
nonlinearity), and we can make this nonlinearity explicit.

\section{Photon counting}

Due to the classical simulatability results of~\cite{Bar01b}, any CV
quantum information process that initiates with finitely-- or
infinitely--squeezed vacua and employs only Clifford group
transformations (phase space displacements, and one-- and two--mode
squeezing), homodyne detection, and classical feedforward, can be
simulated efficiently on a classical computer.  If the cubic phase
gate leads to universal quantum computation with continuous variables,
then it must not satisfy the conditions of this theorem.  (Otherwise,
quantum computation could be simulated efficiently on a classical
computer, which is believed to be impossible.)  The key component of
the cubic phase gate that does not satisfy the conditions of this
theorem is the photon counting PVM; thus, we conclude that the photon
counting PVM cannot be implemented using only Clifford group
transformations and homodyne measurement.

In the following, we show that this transformation cannot be
implemented even with the addition of photodetectors.  We demonstrate
that the photon counting PVM necessarily requires nonlinear evolution.

\subsection{Photon counting using threshold detectors}

The photon counting PVM employed by GKP consists of projections in the
Fock state basis $\{ |n\rangle, n=0,1,\ldots \}$.  For such a
measurement to be performed, one requires photodetectors that can
measure the number of photons in a mode; i.e., distinguish the state
$|n\rangle$ from $|n'\rangle$, $n\neq n'$.  We refer to such a
photodetector as \emph{discriminating}.  However, such discriminating
photodetectors do not yet exist~\cite{Res01}.  All photodetectors in
current use effectively measure whether there are no photons $(n=0)$,
or at least one photon $(n>0)$ in a mode.  (Note that a discriminating
photodetector, on the other hand, must be able to count intrinsically
indistinguishable photons, that is, propagating photons in the same
temporal, spatial and polarization mode~\cite{Blo90}.)  We refer to
existing photodetectors as \emph{threshold detectors}, and a
unit--efficiency detector as an \emph{ideal threshold detector (ITD)}.
The PVM for an ITD is
\begin{equation}
  \label{eq:NonDiscriminatingPVM}
  \bigl\{ \Pi_0 = |0\rangle\langle 0|, \ \Pi_{> 0} =
  \hat{I} - |0\rangle\langle 0| \bigr\} \, ,
\end{equation}
where $\hat{I}$ is the identity operator.  The second projector
$\Pi_{> 0}$ projects onto the infinitely large subspace of states with
one or more photon, due to saturation of the threshold
detector~\cite{Kel64}.

It is possible to use ITDs to distinguish up to a finite number $k$ of
photons by using linear optics to couple the input mode to multiple
ITD modes~\cite{Kok01}.  For example, one could use an array of
beamsplitters to distribute the photons of the input state over $N$
modes such that it is highly unlikely that more than one photon is in
any of the $N$ modes~\cite{Kok01,Kni01}.  ITDs are then used at each
mode, and the probability of undercounting photons is at most
$k(k-1)/2N$.  For small $k$ (as in KLM), the photon counting PVM can
be approximated with high probability by using a sufficiently large
$N$.

In a related fashion, the visible light photon counter
(VLPC)~\cite{Kim99} has been constructed to discriminate between one
and two photons with a high degree of confidence, but the measurement
does not correspond to the photon counting PVM; rather the VLPC is
effective at distributing photons throughout the photosensitive
region, with localized regions acting as threshold devices (as with
standard photodetectors).  In other words, the single--mode input
field is distributed amongst many localised modes in the
photodetector, each region operating as a threshold photodetector.
Consequently the VLPC has much in common with the proposed detection
of photons via arrays of beam splitters to split the signal field (as
discussed above), with an ITD existing at each output port.

Whereas the use of multiple ITDs and linear optics approximates a
discriminating photodetector if the Hilbert space can be truncated as
in qubit--based linear optical quantum computation, this scheme breaks
down for CV quantum computation.  Without \emph{a priori} knowledge of
the maximum number of photons in a mode (for CV, this number is
infinite), one would require an infinite number of auxiliary modes and
ITDs.  (Below, we discuss resource issues even if the Hilbert space is
truncated.)  Thus, the photon counting PVM \emph{cannot} be performed
using linear optics (Clifford group transformations), homodyne
measurement, and a finite number of ITDs.

\subsection{Photon counting with nonlinear evolution}

In order to implement the photon counting PVM, it is illustrative to
employ a model of photodetection based on homodyne measurement and
nonlinear optics (i.e., by employing a Hamiltonian that is cubic or
higher in the photon creation and annihilation operators).  In this
model, a measurement of the phase shift in a probe field allows for a
quantum non--demolition measurement of photon number in the signal
field~\cite{San92}.  Consider the probe field to be a coherent state
with large amplitude.  We interact this probe field with an arbitrary
signal field $|\psi\rangle$ in a Kerr medium with interaction
Hamiltonian
\begin{equation}
  \label{eq:KerrHamiltonian}
  \hat{H}_{\text{int}} = \hbar \chi \hat{N}_{\text{signal}}
  \hat{N}_{\text{probe}} \, ,
\end{equation}
where $\chi$ is proportional to the third--order nonlinear
susceptibility.  After an interaction time $t$, homodyne measurement
is then used to infer the phase shift $\phi$ in the probe field.
Although such a homodyne measurement does not project the probe field
into a phase state, one can in principle project to a state with
arbitrarily small uncertainty $\Delta \phi$ in phase.  A particular
value of $\phi$ can be used to infer the photon number $n=\phi/(\chi
t)$ to the nearest integer value, with corresponding uncertainty
$\Delta n = \Delta \phi/(\chi t)$.  However, the photon number is only
obtained modulo $N = 2\pi/(\chi t)$; the periodicity of the phase does
not give a true photon number measurement~\cite{Mil83}.  This
measurement projects the signal field $|\psi\rangle$ into the subspace
spanned by the number states $|n_j\rangle$, where $n_j = (\phi + 2\pi
j)/\chi$.

To implement the photon counting PVM without issues of periodicity, we
can couple the signal field to a pointer with an unbounded domain,
such as the position~$q$ or momentum~$p$ of a probe.  For example, the
radiation pressure on a mirror is proportional to the flux of
(monochromatic) photons that strike it, and a suitable coupling
Hamiltonian would be~\cite{Pac93}
\begin{equation}
  \label{eq:PositionPointerHamiltonian}
  \hat{H}_{\text{int}} = \lambda \hat{N}_{\text{signal}}
  \hat{q}_{\text{probe}} \, ,
\end{equation}
which is also nonlinear.  For a probe field initially in the momentum
eigenstate $|p=0\rangle$, after an interaction time $t$ a measurement of
momentum $p$ of the probe collapses the probe field into a momentum
eigenstate $|p\rangle$ and thereby the signal field into a number
state with $n = p/(\lambda t)$ (again to nearest integer value).

The resulting photon number measurement in either scheme will carry
with it an error (related to measurement precision, and converting
from continuous to discrete quantities).  GKP require that $\Delta n
\ll n^{1/3}$ for a functioning cubic phase gate; this condition places
limits on the acceptable measurement errors.

Thus, measurement of photon number can be described as a nonlinear
interaction plus homodyne measurement.  This result gives insight into
the reason why such measurements can induce a nonlinear
transformation.  Specifically, this model of photon number measurement
is excluded from the conditions for efficient classical
simulation~\cite{Bar01b}.

\subsection{Truncation of photon number}

Naturally, one should be suspicious of the periodicity involved in the
first measurement scheme (employing the Kerr interaction of
Eq.~(\ref{eq:KerrHamiltonian})) and also of the unbounded nature of
the second scheme (employing
Eq.~(\ref{eq:PositionPointerHamiltonian})).  Any physical realization
of a CV quantum information process must have finite energy, and thus
the Hilbert space can effectively be truncated at some highest energy
with photon number $n_{\text{max}}$.  The issue of energy arises in
the second scheme as well, where the momentum increases without bound
and the energy is not bounded above or below.  Even before infinite
energy becomes an issue, a relativistic description must be applied.
Coupling to momentum instead of the position does not eliminate these
difficulties: the displacement itself must also be bounded by the
physical boundaries of the laboratory.  Truncation of the Hilbert
space is an option that, however, presents challenges~\cite{Buz92};
for example, the system is no longer described by continuous variables
but rather by large--dimensional qudits.

An advantage of a truncated Hilbert space is that the linear optics
scheme involving a finite $N$ ITD modes becomes well--defined.
Implementing these photon counting measurement schemes thus becomes a
matter of resources.  Consider the linear optics measurement scheme
for maximum photon number $n_{\text{max}}$.  The probability of
undercounting $k$ photons is at most $k(k-1)/2N$ for $N$ the number of
ITD modes.  Thus, for a fixed probability of undercounting, the
required number of ITD modes scales as $N \propto n_{\text{max}}^2$,
although detector inefficiencies somewhat complicate the issue.  For
the (nonlinear) photon number measurement scheme involving the Kerr
interaction of Eq.~(\ref{eq:KerrHamiltonian}), however, one does not
require additional modes or detectors so this quadratic scaling does
not apply.  All that is needed for the Kerr interaction scheme is an
increase in phase resolution that behaves as $\Delta\phi \propto
n_{\text{max}}^{-1}$.  Similar resolution arguments apply to the
position--pointer scheme.  Thus, even if one were to consider a
truncated Hilbert space, it may be more practical to employ a
nonlinear measurement scheme rather than a multimode ITD scheme; this
result may be true for the KLM and GKP schemes as well.  For true CV
quantum computation, however, multiple ITD arrays cannot suffice, and
nonlinear evolution is a \emph{necessary} component of photon number
measurement.

\section{Conclusions}

Despite the need for nonlinear evolution for photon counting, the
cubic phase gate has a considerable advantage over the use of
nonlinear interactions directly.  Specifically, this gate can be used
to remove the use of nonlinear materials from the computation and
utilize them only in the preparation of cubic phase states.  In other
words, the photon number measurement can be performed ``offline'', and
the cubic phase states can be viewed as a quantum resource to be
prepared prior to the computation.  This way, the states used in the
the computation need not pass through any optical Kerr nonlinearities
with their high absorption, thus avoiding the loss associated with
using such materials.  Also, if the procedure for producing cubic
phase states possesses noise or other sources of error, imperfect
cubic phase states can be purified to produce a smaller number of
states with higher fidelity~\cite{Got01b}.  Again, an advantage is
that this purification can be done offline and is not part of the
computation.  This concept is similar to the KLM scheme, where the
``difficult'' gates are implemented offline on suitable ancilla states
and then quantum teleported onto the encoded states when needed.  In
our scheme, one simply prepares a sufficient number of cubic phase
states prior to the computation, and the entire process may then occur
using only linear optics and homodyne measurement. A key advantage of
this scheme is that the teleportation can be performed
deterministically.

In summary, we have shown that universal CV quantum computation can be
obtained using linear optics (phase--space displacements and
squeezing), homodyne measurement with classical feed--forward, and a
realization of the photon counting PVM.  We describe the PVM for
current (ideal) photodetectors, and demonstrate that such detectors
cannot be used to implement the photon counting PVM with linear optics
alone.  This photon counting PVM carries with it implicit nonlinear
evolution, and we discuss how it can be implemented in a CV system
using a Kerr interaction (or another nonlinear Hamiltonian) and
homodyne measurement.  The resource requirements of this measurement
scheme compared with using linear optics and current photodetectors
are outlined.  Finally, an advantage of this scheme is its use in the
nonlinear gate of GKP, which removes the nonlinear operations from the
computation and reduces them to ``offline'' preparation of ancilla
states.  These results place the implementation of strong nonlinear CV
quantum gates, and thus universal CV quantum computation, in the realm
of experimental accessibility.

\begin{acknowledgments}
  This project has been supported by an Australian Research Council
  Large Grant.  SDB acknowledges the support of a Macquarie University
  Research Fellowship.  We thank E.~Knill and Y.~Yamamoto for useful
  discussions.
\end{acknowledgments}


\begin{thebibliography}{99}
\bibitem{Sho94} P.\ W.\ Shor, ``Proceedings, 35$^{\rm th}$ Annual
  Symposium on Foundations of Computer Science,'' IEEE Press, Los
  Alamitos, CA (1994).
  
\bibitem{Ste98} A.\ Steane, Rept.\ Prog.\ Phys.\ \textbf{61}, 117
  (1998).
  
\bibitem {Nie00} M.\ A.\ Nielsen and I.\ L.\ Chuang, ``Quantum
  Computation and Quantum Information,'' Cambridge University Press,
  Cambridge (2000).

\bibitem {Chu95} I.\ L.\ Chuang and Y.\ Yamamoto, \pra \textbf{52},
  3489 (1995).
  
\bibitem{Kni01} E.\ Knill, R.\ Laflamme, and G.\ J.\ Milburn, Nature
  \textbf{409}, 46 (2001).
  
\bibitem {Got01b} D.~Gottesman, A.~Kitaev and J.~Preskill, \pra
  \textbf{64}, 012310 (2001).

\bibitem {Llo99} S.\ Lloyd and S.~L.\ Braunstein, \prl \textbf{82},
  1784 (1999).
  
\bibitem {Oza97} M. Ozawa, in ``Quantum Communication, Computing, and
  Measurement,'' eds. O. Hirota, A. S. Holevo and C. M. Caves, Plenum
  Press, New York (1997), p. 233.

\bibitem {Cav91} C.~M.~Caves, C.~Zhu, G.~J.~Milburn, and W.~Schleich,
  \pra \textbf{43}, 3854 (1991).

\bibitem {Bar01b} S.~D.~Bartlett, B.~C.~Sanders, S.~L.~Braunstein and
  K.~Nemoto, \prl \textbf{88}, 097904 (2002).
  
\bibitem {Mil83} G.\ J.\ Milburn and D.\ F.\ Walls, \pra \textbf{28},
  2065 (1983).
  
\bibitem {Res01} K.~J.~Resch, J. S. Lundeen, and A. M. Steinberg, \pra
  \textbf{63}, 020102(R) (2001).
  
\bibitem {Blo90} K.~J.~Blow, R. Loudon, S. J. D. Phoenix, and T. J.
  Shepherd, \pra \textbf{42}, 4102 (1990).
  
\bibitem {Kel64} P. L. Kelley and W. H. Kleiner, Phys. Rev.
  \textbf{136}, A316 (1964).

\bibitem {Kok01} P. Kok and S. L. Braunstein, \pra \textbf{63}, 033812
  (2001).
  
\bibitem {Kim99} J.~Kim, S. Takeuchi, Y. Yamamoto, and H. H. Hogue,
  \apl \textbf{74}, 902 (1999).
  
\bibitem {San92} B.~C.~Sanders and G.~J.~Milburn, \pra \textbf{45},
  1919 (1992).
  
\bibitem {Pac93} A.~F.~Pace, M. J. Collett, and D. F. Walls, \pra
  \textbf{47}, 3173 (1993).
  
\bibitem {Buz92} V.~Bu\v{z}ek, A. D. Wilson--Gordon, P. L. Knight, and
  W.  K. Lai, \pra \textbf{45}, 8079 (1992).

\end{thebibliography}
\end{document}